\documentclass[seceq]{ptptex}

\newcommand{\eq}{\begin{equation}}
\newcommand{\eqx}{\end{equation}}
\newcommand{\eqs}{\begin{equation*}}
\newcommand{\eqsx}{\end{equation*}}
\newcommand{\eqn}{\begin{eqnarray}}
\newcommand{\eqnx}{\end{eqnarray}}
\newcommand{\alg}{\begin{align}}
\newcommand{\algx}{\end{align}}

\newcommand{\f}[2]{\frac{#1}{#2}}

\newcommand{\cor}[1]{\left\langle{#1}\right\rangle}

\newcommand{\al}{\alpha}
\newcommand{\bt}{\beta}

\newcommand{\eps}{\varepsilon}

\newcommand{\nn}{{\cal N}}

\title{
AdS/CFT and the dynamics of quark-gluon plasma%
}

\author{
Romuald A. \textsc{Janik}%
}

\inst{
Institute of Physics,
Jagiellonian University\\
ul. Reymonta 4,
30-059 Krak\'ow,
Poland
}

\abst{
In this talk we first discuss the motivation for using the AdS/CFT
correspondence as a tool for the understanding of real-time dynamics of quark-gluon plasma.
After describing the effective dual degrees of freedom for a strongly coupled gauge theory
system, we review the subsequent theoretical `layers' in the physics of quark-gluon plasma: a uniform plasma at fixed temperature, linearized fluctuations, nonlinear hydrodynamic behaviour and finally some aspects of addressing non-equilibrium physics. The discussion is brief and aimed at nonspecialists.
}

\begin{document}

\maketitle

\section{Introduction}

There are strong indications that the quark-gluon plasma formed in relativistic heavy-ion collisions is a strongly coupled system. This poses a formidable challenge for our understanding of the physics involved.
For static properties, like thermodynamical properties at fixed temperature, we have at our disposal the standard tools of lattice QCD which provide us with information in the nonperturbative regime. These methods have the advantage of working directly with QCD and thus provide quantitative results.
Their drawback is that they work in the Euclidean signature and hence one cannot use them to study real-time properties during the expansion of the plasma system, its approach to equilibrium etc. In principle one can sometimes use these methods and analytically continue to Minkowski signature, but this is very difficult and relies on nontrivial dynamical input (assumptions on spectral functions etc.). Even then one can access only the simplest observables like shear viscosity at fixed temperature. 

An alternative route is of course to use phenomenological models, like hydrodynamics with an equation of state fitted to best describe the experimental data. 
We will be interested here, however, in
a theoretical description `from first principles' where we could understand e.g. how hydrodynamics can be \emph{derived} as well as how one could study also far from 
equilibrium dynamics of the plasma. Thus the main goal of the approach presented here is a qualitative theoretical understanding of the dynamics rather than a quantitative phenomenological description of experimental data.

The use of the AdS/CFT correspondence \cite{adscft} for studying the behaviour of strongly coupled plasma comes from the fact that relatively simple analytical methods capture a lot of the nonperturbative plasma physics, and eventual numerical methods are incomparably simpler than a full fledged lattice 
QCD simulation. Moreover, the AdS/CFT correspondence works equally well in Minkowski as well as in Euclidean signature making it applicable to real-time dynamical problems.
There is one caveat, however, the AdS/CFT correspondence in its current versions cannot be directly applied to QCD. Hence one has to study plasma physics in a \emph{different} theory -- like the $\nn=4$ Super-Yang-Mills theory. 

Despite the fact that QCD and $\nn=4$ SYM are completely different at zero temperature, once we turn on a nonzero temperature, the essential differences like lack of scale or supersymmetry are broken by the nonzero temperature. Both theories are strongly coupled and (in the case of QCD above $T_c$) nonconfining. Therefore one may expect to obtain some \emph{qualitative} insight into the dynamics of plasma from the AdS/CFT correspondence.

However there are also significant differences between these theories. In particular there is no running coupling in $\nn=4$ SYM, that theory has exactly conformal equation of state and no confinement-deconfinement phase transition. Hence an expanding $\nn=4$ plasma cools indefinitely.

Therefore, the applicability of using the simplest $\nn=4$ SYM version of AdS/CFT to model real quark-gluon plasma depends very much on the questions asked.

\section{Effective degrees of freedom at strong coupling}

According to the AdS/CFT correspondence, four-dimensional $\nn=4$ supersymmetric gauge theory is
exactly equivalent to superstring theory in the ten-dimensional $AdS_5 \times S^5$ background. The $AdS_5$ part of it is defined by the metric
\eq
\label{e.ads}
ds^2=\f{\eta_{\mu\nu} dx^\mu dx^\nu+dz^2}{z^2}
\eqx
with $z>0$ and $z=0$ being the boundary of $AdS_5$.

The dual degrees of freedom for $\nn=4$ SYM are strings in $AdS_5 \times S^5$. Different vibrational modes of the string represent various fields/particles. The massless modes include the graviton etc. In addition there is an infinite set of massive modes. This complicated picture simplifies drastically at strong coupling, where all the massive modes become very heavy and thus their dynamics decouples and the dual description reduces just to (super)gravity.

Let us now analyze how time-dependent evolution of strongly coupled plasma is encoded in the dual gravitational description. We expect the plasma system to correspond to a very complex excitation of $AdS_5$ with very many gravitons. Thus it may be better to describe it as a modification of the geometry (\ref{e.ads}):
\eq
\label{e.plasma}
ds^2=\f{g_{\mu\nu}(x^\rho,z) dx^\mu dx^\nu+dz^2}{z^2}
\eqx
This dictionary can be made quite precise, with the energy momentum tensor of the gauge theory plasma system being directly expressible in terms of the geometry (\ref{e.plasma}). Namely $\cor{T_{\mu\nu}(x^\rho)}$ can be recovered from the expansion of $g_{\mu\nu}(x^\rho,z)$ near the boundary $z=0$:
\eq
\label{e.gexp}
g_{\mu\nu}(x^\rho,z)=\eta_{\mu\nu}+z^4 g^{(4)}_{\mu\nu}(x^\rho)+\ldots
\eqx
with the energy-momentum tensor being given by \cite{Skenderis}
\eq
\cor{T_{\mu\nu}(x^\rho)} = \f{N_c^2}{2\pi^2} \cdot g^{(4)}_{\mu\nu}(x^\rho)
\eqx
Applying the dictionary in the opposite direction, we can recover the geometry corresponding to a given spacetime profile of the energy-momentum tensor by solving the \emph{five-dimensional} Einstein's equations
\eq
R_{\al\bt}-\f{1}{2}g^{5D}_{\al\bt} R - 6\, g^{5D}_{\al\bt}=0
\eqx
with the boundary condition (\ref{e.gexp}). For generic choices of $\cor{T_{\mu\nu}(x^\rho)}$ such a geometry will have naked curvature singularities. The possible physical profiles 
will be the ones which lead to a nonsingular geometry \cite{JP1}.

A classical example of this construction is a static, uniform, isotropic plasma system with a constant traceless diagonal energy-momentum tensor. Solving the Einstein's equations with the boundary condition (\ref{e.gexp}) yields the standard AdS planar black hole, which in these coordinates takes the form:
\eq
\label{e.bh}
ds^2=-\f{(1-z^4/z_0^4)^2}{(1+z^4/z_0^4)z^2}\ dt^2
+(1+z^4/z_0^4)\f{dx^2_i}{z^2}+ \f{dz^2}{z^2}
\eqx
where the position of the horizon, $z_0$, is related to the energy density of the plasma configuration $E$ through
\eq
E=\f{3 N_c^2}{2\pi^2 z_0^4}
\eqx
The Hawking temperature and Bekenstein-Hawking entropy are identified with the gauge theory temperature and entropy.

\section{Strongly coupled time-dependent plasma --- linearized theory}

Let us now consider small perturbations of the uniform plasma system described at the end
of the previous section. On the gauge theory side we may expect that the low energy excitations have a hydrodynamic character with the dispersion relation involving shear viscosity in the damping term. There may also be other excitations which are strongly damped but for which there is no phenomenological framework.

On the AdS side of the AdS/CFT correspondence, the excitations will be represented by gravitational perturbations of the 5-dimensional planar black hole (\ref{e.bh}). These perturbations will have to have purely ingoing boundary conditions at the horizon (so-called quasinormal modes in General Relativity). The lowest ones indeed have a dispersion relation characteristic of hydrodynamic modes (with the specific value of the shear viscosity to entropy ratio $\eta/s=1/4\pi$ \cite{shear}) for low momenta \cite{lin1}. However, the Einstein's equations determine this dispersion relation to all orders in $p$, which would be modeled by higher order viscous hydrodynamics \cite{lin2}. Apart from that, there is an infinite tower of highly damped quasi-normal modes, whose damping rate is again uniquely determined by Einstein's equations and corresponds physically to the rate of approach to equilibrium for the given perturbation \cite{horhub}.

Thus Einstein's equations reproduce not only linearized hydrodynamic behaviour with specific values of the transport coefficients, but also all highly damped modes.

Let us note that from the point of view of gravitational physics the fact that we used here the linearized approximation was purely a matter of choice. The dynamical equations governing the dual description of gauge theory system at strong coupling --- five-(or ten-)dimensional Einstein's equations are fully nonlinear. Thus one may use them to
\begin{enumerate}
\item[i)] \emph{derive} hydrodynamic behaviour in the fully nonlinear regime
\item[ii)] address nonequilibrium phenomena which cannot be described, even in principle, by hydrodynamic modes 
\end{enumerate}
We will now briefly review both of these developments.

\section{Strongly coupled time-dependent plasma --- nonlinear hydrodynamic regime}

The first derivation of nonlinear hydrodynamic behaviour from the AdS/CFT correspondence was performed in \citen{JP1}, where  a configuration of plasma invariant under longitudinal boosts (with no dependence on transverse coordinates) was considered. Under these assumptions the whole energy-momentum tensor is uniquely expressed in terms of a single function $\eps(\tau)$ --- the energy density at mid-rapidity. 

In \citen{JP1}, the power $s$ in the late time asymptotics of $\eps(\tau) \sim 1/\tau^s$ was fixed from the AdS/CFT correspondence. Firstly, for each $s$  an asymptotic solution of Einstein's equations was found which described a plasma configuration evolving according to $\eps(\tau) \sim 1/\tau^s$. Secondly, it was found that the only value of $s$ for which the constructed gravitational solution did not have a (naked) curvature singularity was $s=4/3$ leading to the conclusion that the only physically acceptable power-like asymptotics of $\eps(\tau)$ is
\eq
\label{e.perfect}
\eps(\tau) \sim \f{1}{\tau^{\f{4}{3}}}
\eqx
which is exactly the value expected from nonlinear hydrodynamics (Bjorken flow\cite{bjorken}).

Subsequently subleading corrections to (\ref{e.perfect}) were determined \cite{shinvisc,RJvisc,RJMH} consistent \cite{RJvisc} with the value of shear viscosity extracted from the linearized regime\cite{shear}. Moreover second order transport coefficients were found for the first time \cite{RJMH} and the behaviour of $\eps(\tau)$ was determined to be
\eq
\label{e.eps}
\eps(\tau)=\f{1}{\tau^{{\f{4}{3}}}} -{\f{2}{2^{\f{1}{2}} 3^{\f{3}{4}}}} \f{1}{\tau^{{2}}} +
 {\f{1+2\log 2}{12 \sqrt{3}}} \f{1}{\tau^{{\f{8}{3}}}} +\ldots
\eqx

In a further very interesting development the appearance of nonlinear hydrodynamics from gravity was understood without any symmetry assumptions \cite{Minwalla}. On the AdS side, hydrodynamics corresponds to solving the Einstein's equations in a gradient expansion w.r.t. the boundary (i.e. physical gauge theory) coordinates. The expansion is made about a boosted black hole, whose velocity and temperature are promoted to slowly varying functions of the boundary coordinates. 
The requirement of nonsingularity, just as for the boost-invariant case, fixes the integration constants which correspond to transport coefficients of viscous hydrodynamics.

\section{Strongly coupled time-dependent plasma --- far from equilibrium regime} 

Let us emphasize that just as in the case of linearizing Einstein's equations, nothing in their structure forces us to use a gradient expansion. If we treat Einstein's equations exactly (e.g. by solving them numerically) we will be able to describe also far from equilibrium situations where e.g. gradients are large or a process which does not couple to hydrodynamic modes. The simplest example of the first situation is boost-invariant flow but considered in the regime of small proper times. An example of the second situation is the study of the isotropisation of an anisotropic spatially uniform plasma system.

In the latter case the energy-momentum tensor has the diagonal form
\eq
T_{\mu\nu}=
\begin{pmatrix}
		\eps & 0 & 0 & 0 \\
		0 & p_\parallel(t) & 0 & 0 \\
		0 & 0 & p_\perp(t) & 0 \\
		0 & 0 & 0 & p_\perp(t) \\
\end{pmatrix}
\eqx
Here the anisotropy in the longitudinal and transverse pressure cannot be attributed
to flow so the time evolution cannot be described at all even by `all-order' viscous hydrodynamics. However the analysis using Einstein's equations can be applied to this case without any conceptual difficulty albeit technically this is quite involved.
The AdS/CFT approach for studying such a system was proposed in \citen{aniso}. The first numerical
investigation of isotropisation of the plasma relaxing after an anisotropic perturbation of the gauge theory metric was performed in \citen{CY1}.

The second regime where nonequilibrium behaviour of the plasma becomes important is
the small proper time regime of the boost-invariant flow described earlier. As can be seen 
from the expansion (\ref{e.eps}), as we decrease $\tau$ more and more orders of viscous hydrodynamics start to be important and obviously for $\tau \sim 0$, the organization of the result into a hydrodynamic expansion does not make sense. As mentioned before, Einstein's equations do not force one to use a gradient expansion so one can use them also to study this regime. Physically it is clear that the dynamics will depend on the initial conditions and in fact the key question is to understand how the transition to hydrodynamic evolution 
depends on the initial condition.

Currently two approaches to study this problem have been implemented. One is to prepare the
state by a boost-invariant deformation of the gauge theory metric, which has been studied numerically in \citen{CY2}, the other is to keep the gauge theory metric to be Minkowski and to study the evolution starting from some
initial geometry at $\tau=0$. This approach has been pursued analytically in \citen{BHJP}, and
numerical simulations are currently in progress \cite{WIP}.

Hopefully, these investigations will shed light on the fast thermalization time observed 
(indirectly) for quark-gluon plasma produced at RHIC.

In the above discussion we omitted several other topics and developments 
concerning nonequilibrium plasma dynamics. For more detailed discussion and references 
see e.g. the recent reviews \cite{RJlectures} and \cite{Mukund}.

\section{Conclusions}

The AdS/CFT correspondence provides a very powerful tool for studying real-time dynamical
properties of a strongly coupled plasma system in $\nn=4$ SYM theory. By translating the 
physical problem into the dual gravitational language, one can use Einstein's equations as
a unified approach which encompasses both hydrodynamic behaviour in its linearized or fully nonlinear form, as well as fully nonequilibrium processes.

The complexity of Einstein's equations makes it still quite difficult to answer the outstanding
problem of understanding fast thermalization, but the problem seems tractable at least numerically.

Finally it would be very interesting to reconsider some of these problems in a more realistic 
version of the AdS/CFT correspondence which would be closer to real-world QCD. In particular it would be interesting to understand, even qualitatively the impact on the real-time plasma dynamics of the existence of a confinement/deconfinement phase transition, a breaking of scale invariance and eventually perhaps running of the coupling.

\section*{Acknowledgements}
This work is supported in part by the Yukawa International Program for Quark-Hadron Sciences at Yukawa Institute for Theoretical Physics, Kyoto University during the program {\it New Frontiers in QCD 2010}. This work was supported in part by Polish science funds as a research project N N202 105136 (2009-2011) and the Marie Curie ToK KraGeoMP (SPB 189/6.PRUE/2007/7).

%

\end{document}